 \documentclass[epj]{svjour}

\newcommand\epjc[3]  {{Eur.\ Phys.\ J. }{\bf C #1} (#2) #3}

\newcommand\plb[3]   {{Phys.\ Lett.\ }{\bf B #1} (#2) #3}
\newcommand\pr[3]    {{Phys.\ Rev.\ }{\bf #1} (#2) #3}

\newcommand\prl[3]   {{Phys.\ Rev.\ Lett.\ }{\bf #1} (#2) #3}

\newcommand\rmp[3]   {{Rev.\ Mod.\ Phys.\ }{\bf #1} (#2) #3}

\newcommand\jetp[3]  {{Sov.\ Phys.\ JETP\/ }{\bf #1} (#2) #3}

\newcommand\zetf[3]  {{Zh.\ Eksp.\ Teor.\ Fiz.\ }{\bf #1} (#2) #3}

\newcommand\ibid[3]{{ibid.\ }{\bf #1} (#2) #3}

\newcommand{\hepph}[1]{{hep-ph/#1}}

 \usepackage{epsfig,axodraw}

 \title{
  A Gribov equation for the photon Green's function
  }

 \author{K.~Odagiri}

 \institute{
Condensed Matter Physics Group,
Nanoelectronics Research Institute,
National Institute of Advanced Industrial Science and Technology,
Tsukuba Central 2,
1--1--1 Umezono, Tsukuba, Ibaraki 305--8568, Japan
 }

\begin{document}

 \abstract{
  We present a derivation of the Gribov equation for the gluon/photon 
Green's function $D(q)$. Our derivation is based on the second 
derivative of the gauge-invariant quantity $\mathrm{Tr}\ln D(q)$, which 
we interpret as the gauge-boson `self-loop'.
  By considering the higher-order corrections to this quantity, we are 
able to obtain a Gribov equation which sums the logarithmically enhanced 
corrections.
  By solving this equation, we obtain the non-perturbative running 
coupling in both QCD and QED. In the case of QCD, $\alpha_S$ has a 
singularity in the space-like region corresponding to super-criticality 
which is argued to be resolved in Gribov's light-quark confinement 
scenario.
  For the QED coupling in the UV limit, we obtain a $\propto Q^2$ 
behaviour for space-like $Q^2=-q^2$. This implies the decoupling of the 
photon and an NJLVL-type effective theory in the UV limit.
 \PACS{
  {11.10.Lm}{Nonlinear or nonlocal theories and models} \and
  {11.15.Ex}{Spontaneous breaking of gauge symmetries} \and
  {11.15.Tk}{Other nonperturbative techniques} \and
  {12.38.Aw}{General properties of QCD} }
 }

 \date{\today}

 \maketitle

 \section{Introduction}\label{sec_introduction}

  Gribov's programme \cite{gribov,yurireview} for dealing with the 
problem of strongly interacting quarks is based on the picture of 
super-critical binding of light quarks, leading to chiral-symmetry 
breaking and confinement through the reorganization of the quark (the 
Dirac-sea) states in the vacuum \cite{gribovlectures}.

  The central analytical tool used in this approach is the Gribov 
equation.

  In order to derive the Gribov equation, let us consider the 
(all-order) self-energy correction, $\Sigma(q)$, to the quark propagator 
$G(q)$.
  We denote the 4-momentum, $q_\mu$, derivative $\partial/\partial 
q_\mu$ by $\partial_\mu$.
  The raising and lowering of indices is implicit.
  The 4-dimensional Laplacian operator is then defined as 
$\partial^2\equiv\partial_\mu\partial_\mu$, using the Einstein summation 
convention.

  The application of $\partial^2$ to the all-order $\Sigma(q)$ 
yields an infinite series of perturbative diagrams, in which some quark 
or gluon propagator is replaced by its derivative.

  The leading, logarithmically-enhanced, contribution to $\Sigma(q)$ 
comes from phase-space regions with large hierarchies of internal 
momenta.
  Corresponding to this region, the leading contribution to the second 
derivative of $\Sigma(q)$ comes from diagrams in which both derivatives 
act on the same propagator, i.e., $\partial^2 D(q-q')$ where $D(q-q')$ 
is the internal gluon (in this case, bare) propagator carrying 
4-momentum $q-q'$. Neglecting the vacuum polarization contributions, it 
is possible to arrange the internal momenta in such a way that 
derivatives of the quark propagator $G$ do not occur. This simplifies 
things (in the Feynman gauge) since we have the relation:
 \begin{equation}
  \partial^2\frac1{(q-q')^2+i\epsilon}=
  -4\pi^2i\delta^{(4)}(q-q'),
  \label{eqn_laplace_operator}
 \end{equation}
  which removes a loop integration and make it formally equivalent to 
the amplitude for the emission of two zero-momentum gluons.

  The leading contribution to this double emission is obviously 
$\Gamma_\mu G(q)\Gamma_\mu$, i.e., the successive emission of two gluons 
from a quark line.
  The 0-momentum emission vertex is related to the Green's function by 
the Ward--Takahashi identity:
 \begin{equation}
  \Gamma_\mu(q,q,0)=-\partial_\mu G^{-1}(q).
  \label{eqn_ward_takahashi}
 \end{equation}
  Since $\Sigma(q)$ gives the running of $G^{-1}(q)$, the end result of 
this discussion is the following Gribov equation which sums all the 
leading logarithmically enhanced contributions:
 \begin{equation}
  \partial^2 G^{-1}(q)=g
  \partial_\mu G^{-1}(q)G(q)
  \partial_\mu G^{-1}(q)+
  \mathcal{O}(g^2).
  \label{eqn_gribov_a}
 \end{equation}
  The coupling $g$ is $\alpha/\pi$ for QED and $C_F\alpha_S/\pi$ for 
QCD. 
  Because of the removal of the loop integration, this equation is 
local in the momentum space, and benefits from the absence of the 
momentum space integration over the dangerous IR (infra-red) region.

  As mentioned above, the renormalization of the gluon, or the photon in 
QED, propagator has been neglected in this discussion, though it is 
possible to incorporate this effect partly by replacing $g$ in 
eqn.~\ref{eqn_gribov_a} by a running coupling, $g(q)$.

  Leaving aside the problem of the coupled evolution of the quark and 
gluon Green's functions for now, the problem is to write down an 
analogous Gribov equation for the gluon/photon sector.

  In this case, the vertex $\Gamma(q,q',q-q')$ is evidently no longer 
constant at the tree level, and is linear in the momenta. Therefore the 
simple application of the above method, of applying $\partial^2$ to the 
vacuum-polarization operator $\Pi_{\lambda\sigma}$, is insufficient in 
the sense that it yields extra contributions due to the derivatives of 
the vertex.
  Gribov's solution to this problem \cite{gribov} was to employ the 
linear Duffin--Kemmer formalism, in which the interaction is constant, so 
that we can analyze the problem in a similar way to that of the fermion 
sector. However, the resulting equation, being a coupled second-order 
differential equation, is hardly manageable, and furthermore suffers 
from artificial divergences which plague the Duffin--Kemmer formalism 
\cite{yurireview}.

  On the other hand, the analysis based on the Duffin--Kemmer formalism 
is illuminating at least in the sense that it illustrates how to recover 
the Ward--Takahashi identity, eqn.~\ref{eqn_ward_takahashi}, in the 
zero-momentum limit. This problem and the problem of gauge fixing are 
related. These comprise the main difficulty in formulating a Gribov 
equation for the gluon/photon sector.

  On the other hand, for the problem of the UV (ultra-violet) evolution 
of the QED coupling, Gribov postulated 
\cite{gribov,yurireview,gribovqed} a formulation based on the third 
derivative of the polarization operator $\Pi_{\lambda\sigma}$. Three 
differentiations are necessary to remove the UV divergence. In this 
case, Gribov obtained:
 \begin{equation}
  \left(\frac{d^2}{d\xi^2}+2\frac{d}{d\xi}\right)\frac1g\approx
  -\frac16\left(\left[1-2\Gamma_f\right]^2-3\right)^2.
  \label{eqn_gribov_qed}
 \end{equation}
  $\xi=\ln\left|q^2\right|$, and $\Gamma_f$ is the fermion anomalous 
dimension $d\ln Z_f/d\xi$. In the weak-coupling limit, we have 
$\Gamma_q=0$ and so the right-hand side gives $-2/3$. The ordinary RGE 
(renormalization-group equation) evolution is recovered in this case. In 
the strong-coupling limit, $\Gamma_f$ tends to $(1+\sqrt3)/2$, and so 
the right-hand side tends to zero. In this case, the running coupling 
continues to evolve, but more slowly than in the RGE evolution which 
suffers from the problem of the UV Landau pole.

  In the UV limit, the right-hand side of eqn.~\ref{eqn_gribov_qed} 
tends to $-2/g^2$. This leads to a logarithmic evolution, $g\to\xi$. 
Hence the QED (actually $U(1)_Y$) coupling grows large but finite for 
finite $\xi$ and, as postulated by Gribov \cite{gribovtop}, causes the 
formation of super-critical states, which could be identified with the 
Higgs and Goldstone bosons in EWSB (electro-weak symmetry breaking).

  In the derivation of both eqn.~\ref{eqn_gribov_a} and 
\ref{eqn_gribov_qed}, the running-coupling effects due to the 
renormalization of the internal boson lines are neglected. Furthermore, 
in the QED case, the right-hand side of eqn.~\ref{eqn_gribov_qed} is a 
derivative of a phase-space integral, and so the formulation is not 
strictly local.

  The purpose of this paper is to present an approach to deal with these 
problems. Starting from a gauge invariant expression which corresponds 
to the gauge-boson `self-loop', and collecting the 
logarithmically-enhanced terms as in the derivation of 
eqn.~\ref{eqn_gribov_a}, we obtain a Gribov equation for gauge bosons, 
and an evolution equation for the running coupling, in a closed form.

  It is found that the equation can be integrated analytically.
  We analyze the solution for both QCD and QED.

  In QCD, the solution exhibits a branch-cut singularity at space-like 
momentum $Q^2=-q^2=\Lambda^2_\mathrm{QCD}$, indicating vacuum 
instability due to the formation of the critical state.
  The problem of vacuum instability is resolved in Gribov's scenario due 
to the reorganization of the quark states which is best understood using 
the Dirac-sea picture. We argue that this will remove the singularity 
and give rise to a Green's function for the gluon which has the 
analytical properties that are expected for confined particles.

  If, on the other hand, there are no quark states which become 
super-critical, there is a problem as to how to stabilize the vacuum, 
and we believe that this is not possible within the framework of 
ordinary field theory.

  In the case of QED at high energy, we found that the coupling grows 
linearly with $Q^2=-q^2$.
  This corresponds to the decoupling of the photon, and an effective 
high-energy theory described by a contact interaction term, as in the 
NJLVL model \cite{njlvl}.
  This evolution of the coupling is faster than the logarithmic 
behaviour found by Gribov. We show that provided that one includes the 
contribution due to internal photon renormalization in 
eqn.~\ref{eqn_gribov_qed}, the $\propto Q^2$ behaviour of the coupling 
is obtained also in this latter case.

  This paper is organized as follows. We describe the framework of our 
approach in sec.~\ref{sec_framework}. We derive the main equation in 
sec.~\ref{sec_derivation}. Its solution is presented in 
sec.~\ref{sec_solution}. We discuss its application in QCD and QED in 
secs.~\ref{sec_qcd} and \ref{sec_qed} respectively. The conclusions are 
stated at the end.

 \section{General framework}\label{sec_framework}

  Let us begin with the gluon/photon Green's function:
 \begin{equation}
  D(q,\zeta)=Z(q^2)\frac{ -g_{\lambda\sigma}
        +(1-\zeta^{-1})q_\lambda q_\sigma/q^2}{q^2+i\epsilon},
  \label{eqn_gluon_gf}
 \end{equation}
  where $\zeta$ is the gauge-fixing parameter and $Z(q^2)$ is the 
renormalization coefficient. The running coupling, $\alpha(q^2)$, is 
proportional to $Z(q^2)$.

  The central quantity which we adopt in the following discussion is the 
trace of the logarithm of $D(q,\zeta)$, which we denote by $\Xi_\zeta$.
 \begin{equation}
  \Xi_\zeta(q)=\mathrm{Tr}\ln D(q,\zeta).
  \label{eqn_xi_zeta_defn}
 \end{equation}
  Although this quantity will be shown to be essentially gauge 
invariant, we introduced the index $\zeta$ to indicate that $\Xi_\zeta$ 
includes an unphysical polarization contribution.

  We define the logarithm in eqn.~\ref{eqn_xi_zeta_defn} by its Taylor 
series expansion, i.e.:
 \begin{equation}
  \ln\left[\lambda(I+M)\right]=I\ln\lambda+M-\frac{M^2}2+\frac{M^3}3
   -\ldots,
 \end{equation}
  where $\lambda$ is scalar and $M$ is a $d\times d$ matrix. $I$ is the 
$d$-dimensional identity matrix. Applying this to 
eqn.~\ref{eqn_xi_zeta_defn}, we obtain:
 \begin{equation}
  \Xi_\zeta(q)=4\ln\left(\frac{-Z(q^2)}{q^2+i\varepsilon}\right)+
   \ln\zeta^{-1}.
  \label{eqn_xi_zeta_result}
 \end{equation}
  This result is for the 4-dimensional vector boson, but is easily 
generalizable to arbitrary particles. Since the $\zeta$ dependence can 
be absorbed in the definition of $Z(q^2)$ or the scale of $q^2$, 
$\Xi_\zeta$ is essentially gauge invariant.

  $\Xi_\zeta$ corresponds diagrammatically to the self-loop of the gauge 
boson, i.e., just a circle, without the phase-space integration.
  This assignment is natural because when we take its derivative, we 
obtain the trace of a propagator $D$ attached to a $0$-momentum vertex, 
$-\partial D^{-1}$. Explicitly:
 \begin{equation}
  \partial_\mu\Xi_\zeta(q)=\mathrm{Tr}\left[
   -\partial_\mu D^{-1}(q) D(q)\right],
  \label{eqn_xi_amu}
 \end{equation}
  which follows directly from eqn.~\ref{eqn_xi_zeta_defn}.
  We represent this relationship diagrammatically as:
 \begin{equation}
  \begin{picture}(100,60)
   \SetWidth{2}
   \GCirc(15,30){15}{1}
   \LongArrow(40,30)(60,30) \rText(45,20)[][0]{$\partial_\mu$}
   \GCirc(85,30){15}{1} \DashLine(85,45)(85,60){4}
  \end{picture}
  \label{eqn_xi_amu_diagram}
 \end{equation}
  The thick solid line represents $D(q,\zeta)$ and the dashed line 
represents $-\partial D^{-1}(q,\zeta)$.

  We would like to note at this point that the quantity inside the trace 
on the right-hand side of eqn.~\ref{eqn_xi_amu} has analogous form to 
$A_\mu(q)$ introduced in ref.~\cite{gribov}:
 \begin{equation}
  A_\mu(q)=\partial_\mu G^{-1}(q)\ G(q).
  \label{eqn_def_amu}
 \end{equation}
  Since we have:
 \begin{equation}
  \partial_\mu G(q)\equiv -G(q)\partial_\mu G^{-1}(q)G(q)
  \equiv -G(q)A_\mu(q),
 \end{equation}
  the derivative of a propagator is equivalent to its multiplication by 
$-A_\mu(q)$ and is, by the Ward--Takahashi identity, equivalent to the 
insertion of the $0$-momentum photon/gluon emission vertex. It is hence 
appropriate to represent $\Xi_\zeta(q)$ as a self-loop.

  In ref.~\cite{gribov}, $A_\mu(q)$ was introduced to reduce 
eqn.~\ref{eqn_gribov_a} to a first-order equation, namely:
 \begin{equation}
  \partial_\mu A_\mu(q)+(1-g)A_\mu(q)A_\mu(q) = 0.
  \label{eqn_gribov_b}
 \end{equation}
  We reproduce this equation here for comparison with the photon/gluon 
Gribov equation, which we shall write in a similar form later.

  As for the physical interpretation of $\Xi_\zeta$, the logarithm of 
the propagator has the interpretation as the density of states 
\cite{luttingerward}.
 Indeed, we see that the phase-difference:
 \begin{equation}
  \frac1\pi\mathrm{Im}\left[\Xi_\zeta(q_2)-\Xi_\zeta(q_1)\right],
 \end{equation}
  by the Levinson theorem, is the number of states in that phase-space 
interval.
  However, as stated earlier, $\Xi_\zeta$ includes an unphysical 
polarization contribution, and the number of states should in fact be 
$3/4$ of this. This can be achieved by eliminating the scalar 
polarization component in eqn.~\ref{eqn_xi_zeta_defn}:
 \begin{equation}
  \Xi(q)=\mathrm{Tr}P(q)\ln D(q,\zeta)=3\ln\left(
  \frac{-Z(q^2)}{q^2+i\varepsilon}\right).
  \label{eqn_xi_result}
 \end{equation}
  Here, $P$ is the transverse projection operator:
 \begin{equation}
  P_{\lambda\sigma}=g_{\lambda\sigma}-\frac{q_\lambda q_\sigma}{q^2}.
 \end{equation}
  $\Xi(q)$ is gauge invariant.

  We may equally have chosen the derivative of $\Xi$ as the starting 
point of our discussion. Let us define:
 \begin{equation}
  A_\mu(q,\zeta)=\partial_\mu D^{-1}(q,\zeta)D(q,\zeta).
 \end{equation}
  We also define its transverse component:
 \begin{equation}
  A^T_\mu(q)=PA_\mu(q,\zeta)P,
 \end{equation}
  which is gauge invariant. We can then define $\Xi$ as the indefinite 
integral of $A^T$, i.e.:
 \begin{equation}
  \Xi(q)=-\mathrm{Tr}\int^q A^T_\mu(q',\zeta)dq'_\mu.
 \end{equation}
  Note that this is a line integral and not a phase-space volume 
integral which $\Xi(q)$ certainly is not.
  In general, because of the singularities in $D(q,\zeta)$, the value of 
$\Xi$ depends on the path of integration in the complex plane. In 
particular, if we adopt a closed contour for the integration, we have an 
Aharonov--Bohm-type phase corresponding to the `gauge field' 
$\mathrm{Tr}A^T_\mu(q)$.
  The value of the phase is $2i\pi$ times the number of states enclosed 
by the contour.

  Using the above tools, the discussion of the Ward--Takahashi identity 
becomes relatively simple. To see this, let us write the $0$-momentum 
limit of the bare $3$-point vertex using the usual Feynman rules as:
 \begin{equation}
  \Gamma_{\mu,\mathrm{bare}}(q,q,0)=2q_\mu g_{\lambda\sigma}-
    q_\lambda g_{\mu\sigma}-q_\sigma g_{\mu\lambda}.
  \label{eqn_bare_vertex}
 \end{equation}
  We have omitted the colour matrix $f^{ABC}$ for the sake of 
simplicity. This will be reintroduced later, in 
eqns.~\ref{eqn_polarization_lo} and \ref{eqn_dyson_schwinger}, in the 
form of the colour factor $C_A$ inherent in the beta-function 
coefficient $b_0$.
  $D^{-1}$, whose derivative we shall now compare with 
eqn.~\ref{eqn_bare_vertex}, is given by:
 \begin{equation}
  D^{-1}(q,\zeta)=Z^{-1}(q^2)\left[
  -q^2g_{\lambda\sigma}+(1-\zeta)q_\lambda q_\sigma\right].
 \end{equation}
  Taking the bare propagator $D_0^{-1}$, i.e., without considering the 
contribution due to the renormalization $Z(q^2)$, the derivative is 
given by:
 \begin{equation}
  \partial D_0^{-1}(q,\zeta)=
   -2q_\mu g_{\lambda\sigma}+(1-\zeta)\left(
   q_\lambda g_{\mu\sigma}+q_\sigma g_{\mu\lambda}\right).
  \label{eqn_bare_derivative}
 \end{equation}
  The expressions \ref{eqn_bare_vertex} and \ref{eqn_bare_derivative} 
are related by the following tree-level identity:
 \begin{equation}
  \Gamma_\mu(q,q,0)=-\left(\partial_\mu D^{-1}(q,\zeta)-
   \zeta\frac{\partial}{\partial\zeta} D^{-1}(q,\zeta)\right),
 \end{equation}
  exactly as found by Gribov \cite{gribov}. The first term on the 
right-hand side gives the Ward--Takahashi identity, but the second is the 
complication arising from Slavnov--Taylor identity. The latter term 
vanishes only for $\zeta=0$, but the transverse parts are the same.

  Now let us consider operations on $\Xi(q)$. Since this is gauge 
invariant, we may choose the transverse gauge, $\zeta^{-1}=0$. In this 
case, the insertion of a zero-momentum vertex anywhere in the all-order 
diagram for $\Xi(q)$, i.e., including all vacuum-polarization 
contributions, is accompanied by two transverse projectors, and 
therefore the non-transverse part of the vertex do not contribute. We 
hence seem to recover the Ward--Takahashi identity.

  This is not the end of the story, because even though the quantity 
$D(q)\Gamma(q,q,0)D(q)$ is transverse, the quantity 
$\lim_{\zeta^{-1}\to0}\left[D(q)\partial D^{-1}(q)D(q)\right]$ is not 
necessarily so. The trick is to invoke the transversality of $\Xi(q)$. 
For instance, we have:
 \begin{equation}
 -\mathrm{Tr}\left[D(q,\zeta)\partial D^{-1}(q,\zeta)\right]
  = \partial_\mu\Xi_\zeta(q)
  \label{eqn_trace_d_partial_d}
 \end{equation}
  which differs from 
 \begin{equation}
  \mathrm{Tr}\left[D(q,\zeta)\Gamma_\mu(q,q,0)\right]
  = \partial_\mu\Xi(q),
 \end{equation}
  but so long as we make use of $\Xi(q)$, i.e., proceed by taking the 
transverse component of the expression inside the trace in 
eqn.~\ref{eqn_trace_d_partial_d}, the non-transverse component does not 
contribute and the two expressions become identical.
  In summary, the Ward--Takahashi identity:
 \begin{equation}
  \Gamma_\mu(q,q,0)=-\partial_\mu D^{-1}(q,\zeta),
  \label{eqn_ward_takahashi_qcd}
 \end{equation}
  holds provided that we work with gauge-invariant and transverse 
quantities. For transversality, we look for the vanishing of the diagram 
when any gluon in it is assigned scalar polarization, 
$\epsilon_\mu(k)\to k_\mu$.
  This holds for $\Xi(q)$ because of its symmetry: the graph looks the 
same (and transverse) from anywhere in (the perturbative expansion of) 
the diagram.

  Unfortunately, in this discussion, we have lost gauge invariance: the 
transversality argument works only when we utilize a transverse gauge, 
$\zeta^{-1}=0$ with transverse $\Xi(q)$ and $A^T_\mu(q)$.

  A simpler working rule, which reproduces the same result provided that 
we are only interested in products of $D$ and $D^{-1}$, is simply to 
work with the Feynman gauge and impose transversality at the end. To 
illustrate this point, $A_\mu(q,1)$, which is written in the Feynman 
gauge, can be converted to $A^T_\mu(q)$ simply by multiplying on either 
side by the transverse projection operator $P$ as can be easily 
verified:
 \begin{equation}
  A^T_\mu(q)\equiv PA_\mu(q,1)\equiv A_\mu(q,1)P.
 \end{equation}

  We have only discussed the tree-level relationship up to now, but the 
loop corrections decorate both sides of 
eqn.~\ref{eqn_ward_takahashi_qcd} in the same way, and so 
eqn.~\ref{eqn_ward_takahashi_qcd} is an all-order relation.

 \section{Derivation of the equation}\label{sec_derivation}

  In the Feynman gauge, the zeroth-order Gribov equation comes out 
immediately from eqn.~\ref{eqn_laplace_operator}:
 \begin{equation}
  \partial^2 D(q,1)\approx0\qquad (q\ne0).
  \label{eqn_zeroth_order}
 \end{equation}
  In the rest of the discussion, we omit the second argument, $\zeta=1$, 
for simplicity.
  The small contribution to the right-hand side is due to the running of 
the coupling, so that it is $\mathcal{O}(\Gamma)$, where $\Gamma$ is the 
anomalous dimension defined by:
 \begin{equation}
  \Gamma(q^2)=\frac{\partial\ln Z}{\partial\ln q^2}.
 \end{equation}
  This is related to the usual RGE beta function $\beta$ by 
$\Gamma=\beta/\alpha=d\ln\alpha/d\ln q^2$.
  There is a difference of factor $1/2$ between our convention and that 
of ref.~\cite{gribov}.

  Eqn.~\ref{eqn_zeroth_order} can be established in a more gauge 
invariant manner by using $\Xi_\zeta$ introduced in 
eqn.~\ref{eqn_xi_zeta_defn}. Because of its gauge invariance, its 
derivative, $-\mathrm{Tr}A_\mu(q,\zeta)$ is also gauge invariant. We are 
thus justified in using the Feynman gauge. Omitting the trace for 
simplicity, we can then write:
 \begin{equation}
  \partial_\mu A_\mu(q)=\partial\left(\partial D^{-1}(q) D(q)\right)
  =-\partial\left(D^{-1}(q)\partial D(q)\right).
 \end{equation}
  Since:
 \begin{equation}
  \partial D(q)=-D(q)\partial D^{-1}(q)D(q),
 \end{equation}
  and making use of eqn.~\ref{eqn_zeroth_order}, we obtain:
 \begin{equation}
  \partial_\mu A_\mu(q)=A_\mu(q)A_\mu(q)+\mathcal{O}(\Gamma).
  \label{eqn_amu_zeroth}
 \end{equation}

  For the $\mathcal{O}(\Gamma)$ term, since it is due to the running of 
the coupling, we need to analyze the polarization operator 
$\Pi(q^2)=-\Pi_{\mu\mu}(q)/3q^2$. Let us see how this enters our 
framework.

  The expansion of $\Xi(q)$ yields the following:
 \begin{equation}
  \begin{picture}(230,60)
   \SetWidth{2}
   \GCirc(15,30){15}{1}
   \rText(38,30)[][0]{$=$}
   \SetWidth{1}
   \GCirc(67,30){15}{1}
   \rText(89,30)[][0]{$+1$}
   \GCirc(119,30){15}{1}
   \GCirc(119,45){3}{0}
   \rText(141,30)[][0]{$+\frac12$}
   \GCirc(171,30){15}{1}
   \GCirc(171,45){3}{0}
   \GCirc(171,15){3}{0}
   \rText(193,30)[][0]{$+\cdots$}
  \end{picture}
  \label{eqn_pert_expansion_diagram}
 \end{equation}
  The thin lines represent the bare propagator $D_0(q)$, and the blobs 
represent the insertion of the polarization operator. The factors $1/n$ 
are the symmetry factors for the circular symmetry.
  The contribution of the polarization operator, i.e, the sum of the 
contributions from all but the first term of 
eqn.~\ref{eqn_pert_expansion_diagram}, is proportional to:
 \begin{equation}
  \Pi(q^2)+\frac{\Pi^2(q)}2+\frac{\Pi^3(q)}3+\cdots=
  -\ln\left(1-\Pi(q^2)\right).
 \end{equation}
  Including the first term of eqn.~\ref{eqn_pert_expansion_diagram}, we 
thus obtain:
 \begin{equation}
  \Xi(q)=3\ln\left(\frac{-\left(1-\Pi(q^2)\right)^{-1}}{
   q^2+i\varepsilon}\right).
 \end{equation}
  Comparing with eqn.~\ref{eqn_xi_result}, we obtain:
 \begin{equation}
  Z(q^2)\propto\left(1-\Pi(q^2)\right)^{-1}=\alpha(q^2),
  \label{eqn_z_pi_alpha}
 \end{equation}
  which is as expected and consistent with the usual renormalization 
considerations.

  In order to find the sub-leading term in eqn.~\ref{eqn_amu_zeroth}, we 
take the derivative of eqn.~\ref{eqn_pert_expansion_diagram}. Taking the 
first derivative, we have a series of diagrams, approximately half of 
which involving the derivatives of the line and the rest involving the 
derivatives of the blob. The derivatives of the line are already taken 
account of in the leading-order term of eqn.~\ref{eqn_amu_zeroth}, and 
so we are left with the following series of diagrams:
 \begin{equation}
  \begin{picture}(150,60)
   \SetWidth{1}
   \GCirc(15,30){15}{1}
   \GCirc(15,45){3}{0}
   \DashLine(15,45)(15,60){4}
   \rText(35,30)[][0]{$+$}
   \GCirc(63,30){15}{1}
   \GCirc(63,45){3}{0}
   \GCirc(63,15){3}{0}
   \DashLine(63,45)(63,60){4}
   \rText(83,30)[][0]{$+$}
   \GCirc(111,30){15}{1}
   \GCirc(111,45){3}{0}
   \GCirc(111,15){3}{0}
   \GCirc(126,30){3}{0}
   \DashLine(111,45)(111,60){4}
   \rText(140,30)[][0]{$+\cdots$}
  \end{picture}
  \label{eqn_first_derivative_diagram}
 \end{equation}
  The combinatorial factor at each order cancels with the symmetry 
factor present in eqn.~\ref{eqn_pert_expansion_diagram}.

  When we take the next derivative, we again have some derivatives of 
the line, and these are already included in eqn.~\ref{eqn_amu_zeroth}. 
In addition, we have cases in which the two derivatives act on separate 
blobs, such as:
 \begin{equation}
  \begin{picture}(30,60)
   \SetWidth{1}
   \GCirc(15,30){15}{1}
   \GCirc(15,45){3}{0}
   \DashLine(15,45)(15,60){4}
   \GCirc(15,15){3}{0}
   \DashLine(15,15)(15,0){4}
  \end{picture}
  \label{eqn_separated_derivatives_diagram}
 \end{equation}
  These give rise to the renormalization of eqn.~\ref{eqn_amu_zeroth}, 
and therefore are already included. With this consideration, the 
remaining second-derivative terms are:
 \begin{equation}
  \begin{picture}(200,60)
   \SetWidth{1}
   \GCirc(15,30){15}{1}
   \GCirc(15,45){3}{0}
   \DashLine(15,45)(5,55){4}
   \DashLine(15,45)(25,55){4}
   \rText(35,30)[][0]{$+$}
   \GCirc(63,30){15}{1}
   \GCirc(63,45){3}{0}
   \GCirc(63,15){3}{0}
   \DashLine(63,45)(53,55){4}
   \DashLine(63,45)(73,55){4}
   \rText(83,30)[][0]{$+$}
   \GCirc(111,30){15}{1}
   \GCirc(111,45){3}{0}
   \GCirc(111,15){3}{0}
   \GCirc(126,30){3}{0}
   \DashLine(111,45)(101,55){4}
   \DashLine(111,45)(121,55){4}
   \rText(145,30)[][0]{$+\cdots=$}
   \GCirc(185,45){5}{0}
   \DashLine(185,45)(175,55){4}
   \DashLine(185,45)(195,55){4}
   \SetWidth{2}
   \GCirc(185,30){15}{1}
  \end{picture}
  \label{eqn_second_derivative_diagram}
 \end{equation}
  As before, the thick solid line represents the renormalized 
propagator.
  As for the blob, it is essentially $\Pi_{\lambda\sigma}$. However, 
evaluating $\partial^2\Pi_{\lambda\sigma}$ will not yield a sensible 
answer, because there are cancellations between the propagators $\sim 
1/q^2$ and $\Pi_{\lambda\sigma}\sim q^2$ which make the theory 
renormalizable and make eqn.~\ref{eqn_amu_zeroth} valid.

  This $\sim q^2$ factor, in the Feynman gauge and at the, dressed if 
necessary, one-loop level (which is all that is needed), arises from the 
linear momentum dependence of the external vertices, i.e., the vertices 
which are connected with the external propagators.
  Thus we need an operation which extracts only the contributions to 
$\partial^2\Pi_{\lambda\sigma}$ that arises from the derivatives of the 
(dressed) internal lines and not the $\propto q$ linear 
external-momentum dependence of the vertices attached to external 
propagators.
  Let us denote this operation by $\partial_R^2$, where the subscript 
$R$ stands for renormalization.
  With this understanding, we may write the equation in terms of 
$\Pi(q^2)$, as:
 \begin{equation}
  \frac14\mathrm{Tr}\left[
  \partial_\mu A_\mu(q)-A_\mu(q)A_\mu(q)\right]=
  \alpha\partial_R^2\Pi(q^2).
  \label{eqn_amu_allorder}
 \end{equation}
  The left-hand side of this equation is not gauge invariant, and the 
Feynman gauge is implied. It is a trivial matter to convert this into a 
transverse expression, just by substituting $A^T_\mu(q)$ for $A_\mu(q)$ 
and averaging factor $1/3$ for $1/4$.

  It is interesting to explicitly evaluate $\alpha\partial^2\Pi(q^2)$ 
and demonstrate that this does not equal to the left-hand side of 
eqn.~\ref{eqn_amu_allorder}.
  To do so, let us first write out the explicit form of $A_\mu(q)$.
  For general choice of $\zeta$, we have:
 \begin{eqnarray}
  A_\mu(q,\zeta)&=&\partial_\mu D^{-1}(q,\zeta)D(q,\zeta)\nonumber\\
          &=&\frac{2q_\mu}{q^2}\left(1-\Gamma\right)
          g_{\lambda\sigma}+\nonumber\\&&
           \frac{q_\lambda}{q^2}(\zeta-1)P_{\mu\sigma}
             +\frac{q_\sigma}{q^2}(1-\zeta^{-1})P_{\mu\lambda}.
 \end{eqnarray}
  We see that the gauge dependent terms, which are proportional to 
$q_\lambda$ and $q_\sigma$, vanish when either we take the trace of 
$A_\mu$, or in the Feynman gauge. In the Feynman gauge, we have:
 \begin{equation}
  A_\mu(q)=\frac{2q_\mu}{q^2}\left(1-\Gamma\right)
          g_{\lambda\sigma}.
  \label{eqn_amu_feynman_gauge}
 \end{equation}

  It is a simple matter to work out $A^2_\mu(q)$ and $\partial_\mu 
A_\mu(q)$. These are given by:
 \begin{equation}
  A_\mu(q)A_\mu(q)=(1-\Gamma)^2\frac{4g_{\lambda\sigma}}{q^2},
  \label{eqn_a_a}
 \end{equation}
  and
 \begin{equation}
  \partial_\mu A_\mu(q)=-
  \left[\dot \Gamma-(1-\Gamma)\right]\frac{4g_{\lambda\sigma}}{q^2},
  \label{eqn_partial_a}
 \end{equation}
  respectively. Here, $\dot\Gamma$ refers to the derivative by $\ln 
q^2$.
  We may also show that exactly the same relations hold for 
$A^T_\mu(q)$, provided that we replace $g_{\lambda\sigma}$ on the 
right-hand side by $P_{\lambda\sigma}$.

  From eqns.\ \ref{eqn_a_a} and \ref{eqn_partial_a}, we obtain:
 \begin{equation}
  \frac14\mathrm{Tr}\left[
  \partial_\mu A_\mu(q)-A_\mu(q)A_\mu(q)\right]=
  \frac4{q^2}\left[-\dot\Gamma+\Gamma-\Gamma^2\right].
  \label{eqn_delA_minus_Asq}
 \end{equation}
  On the other hand, $\alpha\partial^2\Pi(q^2)$ yields, by using 
eqn.~\ref{eqn_z_pi_alpha}:
 \begin{equation}
  \alpha\partial^2\Pi(q^2)=-Z(q^2)\partial^2Z^{-1}(q^2)=
  \frac4{q^2}\left[\dot\Gamma+\Gamma-\Gamma^2\right],
 \end{equation}
  i.e., similar to eqn.~\ref{eqn_delA_minus_Asq} but with a different 
sign for $\dot\Gamma$.
  Therefore $\partial_R^2\Pi(q^2)$ and $\partial^2\Pi(q^2)$ differ. 

  Now turning our attention to $\partial_R^2\Pi(q^2)$, and temporarily 
adopting the language of perturbation theory with bare propagators, 
$\Pi(q^2)$ has an expansion containing, in general, many gluon lines.

  Following Gribov's argument mentioned in the introduction, the 
logarithmically enhanced contribution to the polarization operator comes 
from regions in the phase space which involve large hierarchy of 
internal momenta. The greatest contribution to $\partial_R^2\Pi(q^2)$ 
comes from terms in which both derivatives act on the same line, and so 
we may apply eqn.~\ref{eqn_laplace_operator}, which removes a momentum 
integration.

  This operation converts the amplitude into that of the emission of two 
zero-momentum gluons, so that the leading contribution will be given by:
 \begin{equation}
  \partial_R^2\Pi(q^2)\propto \mathrm{Tr}\left[A_\mu(q)A_\mu(q)\right]
  +\mathcal{O}(\Gamma),
  \label{eqn_a_a_proportionality}
 \end{equation}
  including all the renormalization correction to the vertices and 
propagators.

  At the one-loop order, we have, independently of the gauge parameter:
 \begin{equation}
  \Pi_1(q)=-4b_0\int\frac{d^4k}{-4\pi^2i}
  \frac1{(k^2+i\epsilon)((q-k)^2+i\epsilon)},
  \label{eqn_polarization_lo}
 \end{equation}
  where the subscript in $\Pi_1$ refers to the perturbative order. $b_0$ 
is the first beta function coefficient, which we define to be positive 
for asymptotically-free theories.
  Applying $\partial_R^2$ to this expression, we obtain:
 \begin{equation}
  \partial_R^2\Pi_1(q)=\partial^2\Pi_1(q)=-4b_0\frac1{q^2+i\epsilon}.
  \label{eqn_double_derivative_pert}
 \end{equation}
  Comparing with the form of eqn.~\ref{eqn_a_a}, this fixes the constant 
of proportionality in eqn.~\ref{eqn_a_a_proportionality} to be $-b_0/4$.

\begin{figure}[ht]{
  \centerline{
  \begin{picture}(100,60)
   \SetWidth{1}
   \Line(5,30)(30,30)
   \put(10,35){\vector(1,0){15}}
   \Text(15,40)[b]{$q$}
   \Line(60,30)(70,30)
   \SetWidth{2}
   \GCirc(45,30){15}{1}
   \put(40,50){\vector(1,0){15}}
   \Text(45,55)[b]{$q-k$}
   \put(40,10){\vector(1,0){15}}
   \Text(45,5)[t]{$k$}
   \GCirc(60,30){3}{0}
   \Text(35,30)[l]{$\lambda$}
   \Text(55,30)[r]{$\lambda$}
  \end{picture}
  }
  \caption{The Dyson--Schwinger-type correction to the one-loop 
gluonic vacuum-polarization operator. The thick lines and the vertex to 
the right are renormalized. The thin line and the vertex to the left are 
bare.
  \label{fig_dyson_schwinger}}}
 \end{figure}
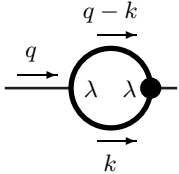

  Now let us work out the same constant of proportionality in an 
all-order approach based on dressed propagators. Inserting (some of the) 
Dyson--Schwinger type all-order corrections, as shown in 
fig.~\ref{fig_dyson_schwinger}, into eqn.~\ref{eqn_polarization_lo}, we 
have:
 \begin{equation}
  \Pi_\mathrm{SD}(q)=
  -\frac{b_0}{3q^2}\mathrm{Tr}\int\frac{d^4k}{-4\pi^2i}
  \Gamma_{\lambda,\mathrm{bare}}D(k)\Gamma_\lambda D(q-k).
  \label{eqn_dyson_schwinger}
 \end{equation}
  As in eqn.~\ref{eqn_bare_vertex}, $\Gamma_\mathrm{bare}$ represents 
the bare vertex.

  As it stands, eqn.~\ref{eqn_dyson_schwinger} is problematic in the 
sense that it is not explicitly gauge invariant, and if it is, we should 
also include the contribution due to the ghost.

  Since we would like to make use of the Ward--Takahashi identity of 
eqn.~\ref{eqn_ward_takahashi_qcd}, it is necessary to work with 
transverse quantities. As discussed in sec.~\ref{sec_framework}, a 
simple way to do this is to work in the Feynman gauge and impose 
transversality in the end.
  It is in fact not even necessary to reintroduce transversality at the 
end, but as a penalty, the spin-averaging factor will be replaced by 
$1/4$.

  We may thus replace $\Gamma_\lambda$ by $-\partial_\lambda D^{-1}$ 
(Feynman gauge), with the understanding that the phase-space region 
giving rise to large logarithms has $q-k\ll q,k$ and hence the 
zero-momentum emission vertex is a good approximation to the full 
vertex.
  In this case, the simplest choice of scale would be $D^{-1}(k)$.
  The integrand in eqn.~\ref{eqn_dyson_schwinger} then becomes:
 \begin{equation}
  \partial_\lambda D_0^{-1}(k)D(k)\partial_\lambda D^{-1}(k) D(q-k).
 \end{equation}

  When we apply $\partial_R^2$ to eqn.~\ref{eqn_dyson_schwinger}, the 
leading renormalization contribution comes from the double derivative, 
$\partial^2$, of $D(q-k)$.
  The application of $\partial^2$ to $D(q-k)$ yields a delta function in 
the approximation that the coupling is constant. If not, we have a 
correction term proportional to $\Gamma$. The first derivative of 
$D(q-k)$ yields:
 \begin{equation}
  \partial_\mu D(q)=\frac{2Z(1-\Gamma)q_\mu}{
  \left((q-k)^2+i\varepsilon\right)^2}g_{\lambda\sigma}.
  \label{eqn_internal_renormalization_effect}
 \end{equation}
  Hence $\partial^2D(q-k)$ is approximately multiplied by factor 
$-Z(1-\Gamma)$ as compared with eqn.~\ref{eqn_laplace_operator}. 
  Factors of $Z$ cancel in the expression for 
$\partial_R^2\Pi_\mathrm{SD}$, which now reads:
 \begin{equation}
  -\frac{b_0(1-\Gamma)}4 \mathrm{Tr}\left[
  \partial_\lambda D^{-1}_0(q)D_0(q)\partial_\lambda D^{-1}(q)D(q)\right].
 \end{equation}
  Let us make the scale choice as $\Gamma(q^2)$.
  The bare propagator $D_0(q)$ and the renormalized propagator $D(q)$ 
differ only by the factor $Z$. Similarly, the difference between 
$\partial_\lambda D^{-1}_0$ and $\partial_\lambda D^{-1}$ is, by 
eqn.~\ref{eqn_amu_feynman_gauge}, $Z^{-1}(1-\Gamma)$. Thus we have:
 \begin{equation}
  \partial_R^2\Pi_\mathrm{SD}(q)=-\frac{b_0}4\mathrm{Tr}\left[
  \partial_\lambda D^{-1}(q)D(q)\partial_\lambda D^{-1}(q)D(q)\right].
 \end{equation}
  We have now shown that by applying Dyson--Schwinger-type corrections 
to the vacuum-polarization graph, we are able to reproduce the form 
which is expected by the logarithmic enhancement argument of Gribov 
which implies eqn.~\ref{eqn_a_a_proportionality} without the need of 
such algebraic manipulations. Although this correspondence between the 
two approaches may seem intuitive and natural, we are not sure about how 
one would go about formulating such a correspondence in the case of the 
fermionic Gribov equation, eqn.~\ref{eqn_gribov_a}.

  In any case, together with eqn.~\ref{eqn_amu_allorder}, we have now 
established the following equation:
 \begin{equation}
  \partial_\mu A_\mu(q)=\left(1-b_0\alpha\right)A_\mu(q)A_\mu(q)+
  \mathcal{O}(\Gamma^2).
  \label{eqn_amu_first}
 \end{equation}
  According to eqn.~\ref{eqn_amu_zeroth}, the term in 
eqn.~\ref{eqn_amu_first} that is proportional to $b_0\alpha$, which 
corrects eqn.~\ref{eqn_amu_zeroth}, is $\mathcal{O}(\Gamma)$. This is 
true, since $-b_0$ is the first expansion coefficient of the beta 
function. We have:
 \begin{equation}
   -b_0\alpha=\Gamma+\mathcal{O}(\Gamma^2).
  \label{eqn_beta_substitution}
 \end{equation}
  Taking the first term, we obtain a rather compact expression with no 
parameter dependence:
 \begin{equation}
  \partial_\mu A_\mu(q)=\left(1+\Gamma(q^2)\right)A_\mu(q)A_\mu(q)+
  \mathcal{O}(\Gamma^2),
  \label{eqn_photon_gribov}
 \end{equation}
  from which we expect $b_0$ to reappear as a constant of integration.
  We shall see in the paragraph following eqn.~\ref{eqn_two_loop_alphas} 
that this reproduces the magnitude of the next-order coefficient, $b'$, 
of the beta function expansion. Therefore the error in 
eqn.~\ref{eqn_photon_gribov} is practically $\mathcal{O}(\Gamma^3)$.
  It is actually difficult to modify eqn.~\ref{eqn_beta_substitution} 
without introducing unphysical fixed points. This point will be 
discussed further in the paragraph leading up to 
eqn.~\ref{eqn_linear_potential_toy_model}

 \section{Solution of the equation}\label{sec_solution}

  Let us now solve our Gribov equation, eqn.~\ref{eqn_photon_gribov}.

  Substituting eqns.\ \ref{eqn_a_a} and \ref{eqn_partial_a} into 
eqn.~\ref{eqn_photon_gribov}, we obtain:
 \begin{equation}
  \dot \Gamma-(1-\Gamma)+(1+\Gamma)(1-\Gamma)^2\equiv
  \dot \Gamma-\Gamma^2(1-\Gamma)=0.
 \end{equation}
  That is,
 \begin{equation}
  \frac{d(\beta/\alpha)}{d\ln{q^2}}=(\beta/\alpha)^2(1-\beta/\alpha).
  \label{eqn_beta_gribov}
 \end{equation}

  $\Gamma=\beta/\alpha$ is positive for QED and negative for QCD. The 
equation is nominally not applicable to the case of QED beyond the 
one-loop order, or to the vacuum polarization due to the quark loop in 
QCD.
  On the other hand, we believe that the formalism is still useful in 
describing the UV behaviour of QED because of the fixed points inherent 
in eqn.~\ref{eqn_beta_gribov}.
  Note that the zeros of the right-hand side of 
eqn.~\ref{eqn_beta_gribov} represent fixed points. The $\Gamma=0$ fixed 
point corresponds to the trivial vacuum whereas $\Gamma=1$ corresponds 
to the QED UV fixed point. In this limit, the running of the QED 
coupling cancels the $1/q^2$ propagator factor, and so the photon 
decouples.

  Even though it is not obvious that the photon should necessarily 
decouple in the UV limit, a limit in which the photon decouples is 
almost necessarily a fixed point of the theory, since in this limit 
there is no longer photon propagation and hence no further evolution of 
the photon propagator. In this regard, the $\Gamma=1$ fixed point of 
eqn.~\ref{eqn_beta_gribov} seems physical.

  On the other hand, the equation will certainly break down when 
discussing effects due to fermion masses or fermionic super-critical 
state formation.
  Related to this point, we do not expect eqn.~\ref{eqn_beta_gribov} to 
be valid when the expansion parameter, i.e., in this case, $\Gamma$, is 
large.

  In the case of UV QED, we think that the evolution is barely 
permissible because of the presence of the physical fixed point, but in 
the case of IR QCD, as there is no fixed point for negative $\Gamma$, 
$\Gamma$ diverges at, as we shall show, $\Lambda_\mathrm{QCD}$.
  In this case, we do not believe that the equation is quantitatively 
correct. However, the behaviour of the solution is, we believe, 
nevertheless physical and, in any case, Gribov's super-critical state 
formation occurs before this singularity. A measure of the quantitative 
accuracy of the equation is provided by the comparison with the 
perturbative beta-function expansion, and this will be presented in the 
next section.

  We would like to note, to avoid confusion, that we also expect the 
$\Gamma=1$ fixed point of eqn.~\ref{eqn_beta_gribov} to be correct in 
the case of QCD because of the decoupling behaviour which it represents. 
  However, the limit indicated by this fixed point does not arise in 
asymptotically-free theories.

  Before proceeding, we note that, had we chosen a different expression 
for eqn.~\ref{eqn_beta_substitution}, we would, in general, end up with 
extra fixed points with unphysical power-like behaviour of the coupling. 
  Because of this, it is difficult to introduce a simple higher-order 
modification to eqn.~\ref{eqn_beta_substitution} without affecting its 
physical behaviour.
  On the other hand, if there arises a need to create a toy model for 
the running coupling with some particular power-like fixed-point 
behaviour, it is easy to artificially modify 
eqn.~\ref{eqn_beta_substitution} to serve this purpose.
  For instance, one may like to introduce an $\alpha_S$ which is finite 
in the space-like region. A possibility would then be the substitution:
 \begin{equation}
  -b_0\alpha=\Gamma-\frac{\Gamma^3}{1-\Gamma}.
  \label{eqn_linear_potential_toy_model}
 \end{equation}
  In this case, corresponding to eqn.~\ref{eqn_beta_gribov}, we obtain:
 \begin{equation}
  \dot\Gamma=\Gamma^2(1-\Gamma^2),
 \end{equation}
  which is self-dual under $\ln q^2\leftrightarrow-\ln q^2$ and 
$\Gamma\leftrightarrow-\Gamma$. In the low-energy limit of QCD, this 
gives us a $\alpha\propto1/q^2$ behaviour, i.e., a simple toy model for 
the long-distance linear potential.

  Resuming our discussion of eqn.~\ref{eqn_beta_gribov}, let us first 
confirm that it leads to the ordinary result for the running coupling 
when $\Gamma$ is small. We would like to calculate the evolution in the 
space-like region, i.e., for positive $Q^2=-q^2$. Omitting the 
sub-leading term, we obtain:
 \begin{equation}
  \beta/\alpha=\left[-\ln Q^2+\mathrm{const.}\right]^{-1}.
 \end{equation}
  The constant on the right-hand side is $\ln\Lambda^2$.
  The left-hand side is the logarithmic derivative of $\ln\alpha$, so 
we obtain:
 \begin{eqnarray}
  \ln\alpha&=&\int \left[\ln(\Lambda^2/Q^2)\right]^{-1} d\ln Q^2
  \nonumber\\&=& -\ln(\ln(\Lambda^2/Q^2))+\mathrm{const.}
 \end{eqnarray}
  This second integration constant gives the leading-order beta-function 
coefficient $b_0$, so we finally obtain:
 \begin{equation}
  \alpha(Q^2)=\frac1{b_0\ln(Q^2/\Lambda^2)}.
  \label{eqn_ordinary_running}
 \end{equation}
  As before, $b_0$ is, in our convention, positive for QCD and negative 
for QED.

  More generally, the integration of eqn.~\ref{eqn_beta_gribov}
yields:
 \begin{equation}
  \ln(\Lambda^2/Q^2)=\Gamma^{-1}+\ln\left|\Gamma^{-1}-1\right|.
  \label{eqn_logq_inverse_Gamma}
 \end{equation}
  The modulus is a shorthand for writing $\Gamma^{-1}-1$ for QED and 
$1-\Gamma^{-1}$ for QCD.
  Even though it is not possible to invert this equation, there is a 
trick to integrate it. We have:
 \begin{equation}
  \log\alpha=\int\Gamma\, d\ln Q^2
  =\int\Gamma\frac{d\ln Q^2}{d\Gamma}d\Gamma
  =\int\frac{\Gamma}{\dot\Gamma}d\Gamma.
 \end{equation}
  Then by the use of eqn.~\ref{eqn_beta_gribov}, we obtain a relation 
between $\alpha$ and $\Gamma$:
 \begin{equation}
  \log\alpha=\int\frac{d\Gamma}{\Gamma(1-\Gamma)}
  =\log\left(\frac{-b_0^{-1}\Gamma}{1-\Gamma}\right).
 \end{equation}
  $b_0^{-1}$ is the constant of integration. Thus we have:
 \begin{equation}
  \left(b_0\alpha\right)^{-1}=1-\Gamma^{-1},
  \label{eqn_alpha_gamma}
 \end{equation}
  which implies the following beta-function expansion:
 \begin{equation}
  \beta=\alpha\Gamma=
  -b_0\alpha^2\left(1+b_0\alpha+b_0^2\alpha^2+\cdots\right).
  \label{eqn_effective_beta_series}
 \end{equation}
  We shall compare this with the QCD beta-function coefficients in the 
paragraph following eqn.~\ref{eqn_two_loop_alphas}.
  We now substitute eqn.~\ref{eqn_alpha_gamma} in 
eqn.~\ref{eqn_logq_inverse_Gamma} to obtain:
 \begin{equation}
  \ln(Q^2/\Lambda_e^2)=(b_0\alpha)^{-1}+\ln|b_0|\alpha,
  \label{eqn_gribov_soln_log}
 \end{equation}
  or:
 \begin{equation}
  Q^2/\Lambda_e^2=|b_0|\alpha\exp\left(1/b_0\alpha\right).
  \label{eqn_gribov_soln_exp}
 \end{equation}
  We have defined $\Lambda_e=\Lambda e^{-1/2}$. In the following, let us 
discuss the properties of this solution.

 \section{The case of QCD}\label{sec_qcd}

  Eqn.~\ref{eqn_beta_gribov} and its solution, eqns.\ 
\ref{eqn_gribov_soln_log} and \ref{eqn_gribov_soln_exp}, are general 
both to QED and QCD.
  However, eqn.~\ref{eqn_beta_gribov} has no zeros for negative 
$\Gamma$, and so its solution becomes non-analytical at 
$Q^2=\Lambda^2=\Lambda_\mathrm{QCD}^2$. At this point, $b_0\alpha$ 
reaches 1.

  To see the behaviour of eqn.~\ref{eqn_gribov_soln_exp} below 
$\Lambda_\mathrm{QCD}$, let us write:
 \begin{equation}
  b_0\alpha=ae^{i\phi}.
 \end{equation}
  Then by taking the imaginary part of eqn.~\ref{eqn_gribov_soln_exp}, 
we obtain
 \begin{equation}
  a\phi=\sin\phi.
  \label{eqn_a_phi_imag}
 \end{equation}
  At $Q^2=\Lambda^2$, $a=1$ and $\phi=0$. Below $Q^2=\Lambda^2$, 
depending on whether we move the singularity above or below the real 
axis, $\phi$ becomes positive or negative. Adopting positive $\phi$, $a$ 
gradually decreases, and by the form of eqn.~\ref{eqn_a_phi_imag}, 
$\phi$ gradually increases. In the limit $Q^2\to0$, $a$ vanishes, and we 
end up with $\phi=\pi$, i.e., $\alpha$ becomes negative. This rotation 
of phase implies that there is one gluonic super-critical state between 
$Q^2=0$ and $Q^2=\Lambda$, with negative mass \cite{gribovlectures}.

\begin{figure}[ht]{
 \centerline{\epsfig{file=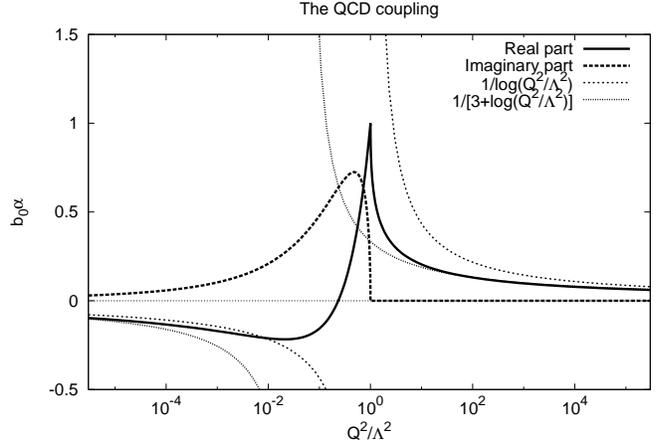, width=9cm}}
 \caption{The QCD running coupling calculated using the Gribov equation,
compared against the leading-order perturbative behaviour, and the
same shifted by $\Lambda^2\to \Lambda^2e^{-2}$.
 \label{fig_qcd_coupling}}}\end{figure}

  Taking the real part of eqn.~\ref{eqn_gribov_soln_exp}, we obtain the 
other constraint:
 \begin{equation}
  \frac{\Lambda_e^2}{Q^2}=\frac\phi{\sin\phi}\exp\left(
  -\frac\phi{\tan\phi}\right).
  \label{eqn_a_phi_real}
 \end{equation}
  We can now make a plot of $b_0\alpha$ against $Q^2/\Lambda$ for both 
real and imaginary regions. In the real region, we use 
eqn.~\ref{eqn_gribov_soln_exp} to evaluate $Q^2/\Lambda$ as a function 
of $b_0\alpha$, whereas in the imaginary region, we make use of eqns.\ 
\ref{eqn_a_phi_imag} and \ref{eqn_a_phi_real} to evaluate $a$ and 
$\Lambda_e^2/Q^2$ as a function of $\phi$. Fig.~\ref{fig_qcd_coupling} 
shows the plot obtained in this way.

  Above $Q^2=\Lambda^2$, we see that there is a modification to the 
one-loop perturbative evolution, which persists up to considerably high 
energy, but most of the modification can be absorbed by a shift in 
$\Lambda_\mathrm{QCD}^2$, found numerically to be about $e^2\approx7$. 
This implies that the measured $\Lambda_\mathrm{QCD}$ is about three 
times smaller than the true $\Lambda_\mathrm{QCD}$ that is obtained in 
the high-energy limit.

  To study this effect, let us consider the iterative inversion of 
eqn.~\ref{eqn_gribov_soln_log}:
 \begin{equation}
  (b_0\alpha)^{-1}=
   \log\left(\frac{Q^2}{\Lambda_e^2}
    \log\left(\frac{Q^2}{\Lambda_e^2}
     \log\left(\frac{Q^2}{\Lambda_e^2}
      \log\Bigl(\cdots\Bigr)
   \right)\right)\right).
 \end{equation}
  At the first level of truncation we obtain 
$\Lambda^2\to\Lambda^2b_0\alpha_0$ where $\alpha_0$ is the value of 
$\alpha$ at some relevant scale, and so on.

  In fact, such a shift in $\Lambda_\mathrm{QCD}$ is known to be present 
already at the perturbative level. One expression for $\alpha_S$ at the 
two-loop order reads \cite{two_loop_alphas}:
 \begin{equation}
  \alpha_S^{-1}(Q^2)+
  b'\ln\left(\frac{b'\alpha_S(Q^2)}{1+b'\alpha_S(Q^2)}\right)
  =b_0\ln(Q^2/\Lambda^2),
  \label{eqn_two_loop_alphas}
 \end{equation}
  where $b'$ is the ratio of the first and second beta-function 
coefficients:
 \begin{equation}
  \beta_\mathrm{perturbative}=
  -b_0\alpha^2(1+b'\alpha+b''\alpha^2+\cdots).
 \end{equation}
  We see that eqn.~\ref{eqn_two_loop_alphas} has almost the same form as 
eqn.~\ref{eqn_gribov_soln_log}. Since the presence of an extra $\alpha$ 
inside the denominator of the second term is a higher-order effect, and 
so is the choice of $\Lambda$, the two equations differ only by the 
difference between $b_0$ and $b'$, as can be inferred from 
eqn.~\ref{eqn_effective_beta_series}. In the real-world QCD, this 
difference is given by:
 \begin{equation}
  \frac{b'}{b_0}=\frac{6(153-19n_f)}{(33-2n_f)^2}=0.790\cdots,
 \end{equation}
  for $n_f=3$. On the other hand, the large-$N_C$, or $n_f=0$, limit of 
the same quantity is $0.843\cdots$. In either case, it is reasonably 
close to unity.

 \begin{table}[ht]
 \begin{center}
 \begin{tabular}{|c|c|c|c|}
 \hline
  & $n_f=3$ & $n_f=0$ & $N_C=\infty$ \\
 \hline
 $\beta_0=4\pi b_0$ & $9$ & $11$ & $11(N_C/3)$ \\
 $\beta_1=(4\pi)^2 b_0b'$ & $64$ & $102$ & $102(N_C/3)^2$ \\
 $\beta_2=(4\pi)^3 b_0b''$ & $3863/6$ & $2859/2$ & $2859(N_C/3)^3/2$ \\
 $\beta_3=(4\pi)^4 b_0b'''$ & $12090.4$ & $29243.0$ & $25554.8(N_C/3)^4$ \\
 \hline
 $\beta_1/\beta_0$ & $7.11$ & $9.27$ & $9.27(N_C/3)$ \\
 $\beta_2/\beta_1$ & $10.06$ & $14.00$ & $14.00(N_C/3)$ \\
 $\beta_3/\beta_2$ & $18.78$ & $20.47$ & $17.89(N_C/3)$ \\
 \hline
 $(\beta_2/\beta_1)/(\beta_1/\beta_0)$ & $1.415$ & $1.510$ & $1.510$ \\
 $(\beta_3/\beta_2)/(\beta_2/\beta_1)$ & $1.867$ & $1.462$ & $1.277$ \\
 \hline
 \end{tabular}
 \end{center}
 \caption{The beta function coefficients and their ratios. The 
three-loop \cite{three_loop_alphas} and four-loop 
\cite{four_loop_alphas} contributions are calculated in the 
$\overline{\mathrm{MS}}$ scheme.
 \label{tab_beta_coefficients_and_ratios}}
 \end{table}

  To analyze the higher-order contributions 
\cite{three_loop_alphas,four_loop_alphas}, let us introduce an 
alternative and common notation for the beta-function coefficients: 
 \begin{equation}
  \beta/\alpha=-\sum_{n=1}^\infty\beta_{n-1}(\alpha/4\pi)^n.
 \end{equation}
  These coefficients are shown in 
tab.~\ref{tab_beta_coefficients_and_ratios} for the three cases: three 
light-quark flavours, zero flavours, and for $N_C=\infty$ which implies 
zero flavours. The zero-flavour case starts to differ from the 
$N_C=\infty$ case only at the four-loop order, where the non-planar 
contributions which are absent up to the three-loop order arise 
\cite{four_loop_alphas}.
  $\beta_2$ and $\beta_3$ are dependent on the renormalization scheme, 
and the numbers quoted in tab.~\ref{tab_beta_coefficients_and_ratios} 
correspond to the $\overline{\mathrm{MS}}$ scheme. The choice of the 
renormalization scheme will affect our discussion here, at least in 
principle.

  One notices that even though the ratio between two successive 
coefficients, $\beta_n/\beta_{n-1}$, deviates away from $\beta_0$ at 
higher orders, it seems to tend to a constant, when the number of 
flavours is zero.
  Of course, it is dangerous to draw any conclusions through knowing 
only these four coefficients, and our knowledge about the higher order 
coefficients is limited, but let us proceed with this tentative 
discussion for now.

 To quantify this statement about the ratio of two successive 
coefficients tending to a constant, we have also tabulated the ratio:
 \begin{equation}
  (\beta_{n+1}/\beta_n)/(\beta_n/\beta_{n-1})=
  \beta_{n+1}\beta_{n-1}/\beta_n^2,
 \end{equation}
  in tab.~\ref{tab_beta_coefficients_and_ratios}.
  For the beta-function expansion to have a finite radius of 
convergence, by d'Alembert's ratio test, it is necessary, though not 
sufficient, for this ratio to tend to one.
  The point at which the beta-function expansion first diverges is the 
point at which super-critical behaviour arises. We expect the ratio to 
be positive, since its being negative would imply super-criticality for 
negative $\alpha_S$, that is, for repulsive strong interaction, and this 
is unphysical.

  The values of $\beta_{n+1}/\beta_n$ being different from $\beta_0$ 
implies that eqn.~\ref{eqn_effective_beta_series} is obviously not 
literally correct, but is an approximation in the sense that up to the 
three-loop order, these ratios are actually close to $\beta_0$, even 
with finite (small) number of flavours.
  This gives partial assurance about the validity of our approach.

  At higher order than three loops, in our opinion, the validity of the 
approach lies not in the numerical accuracy of the Gribov-equation 
evolution but rather in the physics it describes, namely 
super-criticality at finite $\alpha_S$.

  As for the formal presence of the $\Gamma=1$ fixed point in 
eqn.~\ref{eqn_beta_gribov}, this means that in the limit of large 
$\alpha_S$, when the $\beta$-function is analytically continued beyond 
its formal radius of convergence, we expect $\beta/\alpha$ to tend to 
$1$.

  Let us return to the discussion of the branch-point singularity 
mentioned at the beginning of this section. The formation of the gluonic 
super-critical state, indicated by this singularity, would make the 
vacuum unstable, since there is energy gain in, for example, the pair 
production of such states from the vacuum. As far as we can see, there 
is no way to stop or saturate this decay, either statistical or 
dynamical, and so pure QCD has to be, in our opinion, inconsistent.

  However, the situation becomes different when we have light quarks.
  According to Gribov, the bound states of a pair of light quarks 
becomes super-critical when the coupling exceeds the critical coupling 
$\alpha_c$ given by:
 \begin{equation}
  \frac{\alpha_c}\pi C_F=1-\sqrt\frac23=.183\cdots.
  \label{eqn_critical_alpha_value}
 \end{equation}
  This is smaller than the gluonic branch point which occurs at 
$b_0\alpha=1$:
 \begin{equation}
  \frac{\alpha}\pi C_F\Bigr|_{b_0\alpha=1}=
  \frac{16}{33-2n_f}=0.592\cdots.
  \label{eqn_gluonic_alpha_value}
 \end{equation}
  As before, we have taken $n_f=3$. We note that, potentially, the 
effective coupling which appears in the Gribov equation of 
eqn.~\ref{eqn_gribov_a} should be corrected by $(1-\Gamma)$, which 
arises from eqn.~\ref{eqn_internal_renormalization_effect}. This is a 
small effect for the value of $\alpha_S$ given by 
eqn.~\ref{eqn_critical_alpha_value}, but it reduces the critical 
coupling. In addition, the actual values of the beta function 
coefficients, or the ratios of them, tabulated in 
tab.~\ref{tab_beta_coefficients_and_ratios} imply that the gluonic 
super-criticality occurs at much lower values of $\alpha_S$, probably a 
half or so, than is implied by $b_0\alpha=1$. Explicitly, by the ratio 
test for the convergence of the beta function expansion, gluon 
super-criticality occurs at $\alpha_S=\alpha_c^\mathrm{glu}$ given by:
 \begin{equation}
  \alpha_c^\mathrm{glu}/4\pi=
  \lim_{N\to\infty}\frac{\beta_N}{\beta_{N+1}}.
  \label{eqn_alphac_beta}
 \end{equation}
  It is not clear per se whether this limit is actually finite, or that 
it is renormalization-scheme independent. However, our work suggests 
that it is at most finite and is probably non-zero. If it is zero, the 
implication would be that there is some form of super-criticality at any 
value of the coupling, and this appears unlikely to us.

  Returning to the discussion of the presence of the light quarks, since 
it is a non-trivial matter to consider the coupled evolution of the 
gluon and quark Green's functions, let us assume that Gribov's argument 
leading to the formation of light-quark super-critical states is 
essentially unmodified by the running of $\alpha_S$. Obviously this is 
questionable if the singularity at $Q^2=\Lambda^2$ remains, but we 
believe that this is not the case.

  Firstly, due the formation of the light-quark super-critical state, 
the quark Green's function becomes complex, and this corresponds to the 
decay of the light quarks by emitting the super-critical state, or more 
strictly the decay of the vacuum. When this occurs, the gluon Green's 
function also becomes complex, indicating the decay of the gluon by 
emitting the super-critical state through the quarks. This would move 
the singularity off the real axis, and the evolution of $\alpha_S$ will 
continue to $Q^2=0$ without any further singularities.

 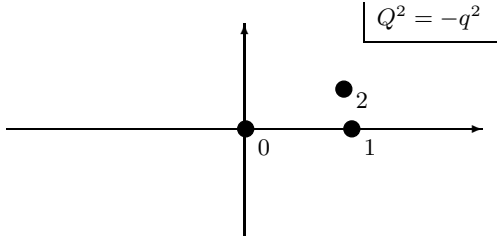
\begin{figure}[ht]{
 \setlength{\unitlength}{1pt}
 \begin{center}
 \begin{picture}(200,100)
 \put(100,10){\vector(0,1){80}}
 \put(10,50){\vector(1,0){180}}
 \put(150,90){\parbox[c]{2cm}{$Q^2=-q^2$}}
 \put(145,83){\line(0,1){15}}
 \put(145,83){\line(1,0){50}}
 \GCirc(100,50){3}{0}
 \Text(105,40)[bl]{$0$}
 \GCirc(140,50){3}{0}
 \Text(145,40)[bl]{$1$}
 \GCirc(137,65){3}{0}
 \Text(142,58)[bl]{$2$}
 \end{picture}
 \end{center}
 \caption{The singularities of the gluon Green's function. The 
branch-point at $Q^2=0$ is labelled $0$, that at $\Lambda^2$ is labelled 
$1$. $2$ corresponds to $1$ moved off axis due to the decay into 
light-quark super-critical states via light quarks.
 \label{fig_singularity_diagram}}}\end{figure}

  This situation is shown in fig.~\ref{fig_singularity_diagram}. In 
one-loop perturbative QCD, there is a branch-point at 
$q^2+i\varepsilon=0$, as well as a simple pole at 
$Q^2=-q^2=\Lambda_\mathrm{QCD}^2$. However, the non-perturbative effects 
inherent in eqn.~\ref{eqn_photon_gribov} tame this and make it a branch 
point, corresponding to gluonic super-critical singularity. The decay 
into light-quark super-critical states move this singularity off the 
real axis.

  This is when the negative-energy super-critical states are not 
completely filled. When they are filled, these decays become forbidden, 
and so $\alpha_S$ becomes real again. When this occurs, the singularity 
near $Q^2=\Lambda_\mathrm{QCD}^2$ should disappear, but there will be a 
new branch cut starting at $q^2-i\varepsilon=0$ which, as in 
ref.~\cite{gribov}, arises as a result of what is best described as the 
reorganization of the Dirac sea.

  To describe the evolution of the Green's functions with the new Dirac 
sea, according to the approach adopted in ref.~\cite{gribov}, it is 
sufficient to modify the evolution equations by including the 
contribution of the Goldstone boson (the pion). Since the Goldstone 
boson only couples to the gluon through quarks, the inclusion of the 
Goldstone boson contribution would be through the light-quark loops in 
the vacuum polarization operator.

  As stated above, the coupled evolution of quark and gluon Green's 
function is a non-trivial matter, but so long as the modification is 
local in the momentum space as indicated by ref.~\cite{gribov}, the IR 
behaviour of $\alpha_S(Q^2)$ is governed by eqn.~\ref{eqn_beta_gribov}. 
  Even if the locality does not hold, in general, we expect the quark 
contribution to the running of the QCD coupling to be small. Here it 
causes a large effect only indirectly, through (the forbidding of) the 
decay of the gluon.

  From the locality of the correction due to the Goldstone boson and the 
analyticity of $\alpha_S(Q^2)$, we may deduce the following.

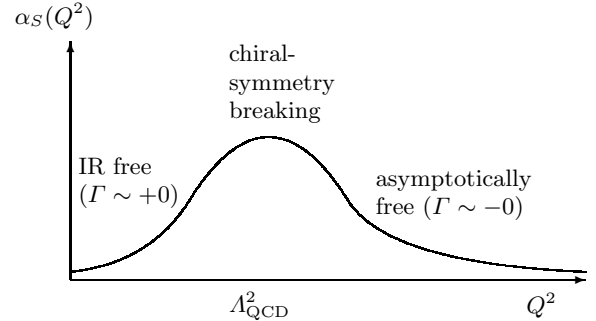
\begin{figure}[ht]{
 \setlength{\unitlength}{1.5pt}
 \begin{center}
 \begin{picture}(150,80)
 \put(20,10){\vector(0,1){60}}
 \put(20,10){\vector(1,0){130}}
 \qbezier(20,12)(40,14)(50,30)
 \qbezier(50,30)(60,46)(70,46)
 \qbezier(70,46)(80,46)(90,30)
 \qbezier(90,30)(100,14)(150,12)
 \put(22,33){\parbox[c]{2.3cm}{IR free\\($\Gamma\sim+0$)}}
 \put(97,30){\parbox[c]{2.3cm}{asymptotically\\free ($\Gamma\sim-0$)}}
 \put(60,58){\parbox[c]{2.3cm}{chiral-\\symmetry\\breaking}}
 \put(60,2){\parbox[c]{2cm}{$\Lambda^2_{\mathrm{QCD}}$}}
 \put(135,2){\parbox[c]{2cm}{$Q^2$}}
 \put(6,75){\parbox[c]{2cm}{$\alpha_S(Q^2)$}}
 \end{picture}
 \end{center}
 \caption{The behaviour of the QCD coupling at low energy.
 \label{fig_qcd_schematic}}}\end{figure}

  If $\Gamma$ remains negative below $Q^2=\Lambda_\mathrm{QCD}^2$, a 
branch-point singularity in $\alpha_S(Q^2)$ again develops at finite 
$Q^2$. Since this would contradict analyticity, $\Gamma$ must change the 
sign. It must also remain below $\Gamma=1$, because otherwise it again 
develops a singularity when $\Gamma$ diverges. As mentioned at the 
beginning of sec.~\ref{sec_solution}, we expect that the $\Gamma=1$ 
fixed point is correct even in asymptotically-free theories in which 
this UV fixed-point limit does not arise.
  We then have an IR-free theory in the IR limit. 
Fig.~\ref{fig_qcd_schematic} illustrates this behaviour.

  Since, as we mentioned above, we do not expect the quark (dynamical) 
contribution to the running of the QCD coupling to be large, the 
evolution in the IR-free region is almost entirely due to the gluons. 
  Thus it may seem strange that the gluons screen the coupling here 
rather than yield the usual anti-screening behaviour.
  However, this is as expected, because in the low-energy effective 
theory, as a result of the forbidden decay to the super-critical states 
(or in other words, the decay into the new states created by Dirac-sea 
reorganization), there is a new cut with the branch point at 
$q^2-i\varepsilon=0$. This means that the Wick rotation has to be 
performed in the opposite direction to the usual one, and so the 
direction of the running is inverted.

  This scenario is as discussed by Gribov \cite{gribov,gribovlectures}. 
The new cuts, which apparently violate causality, correspond to the 
presence of the positive-energy super-critical states of negative-energy 
quarks. Causality is not violated, but it appears as if positive-energy 
quarks are travelling backwards in time.

  An IR-free gluon is decoupled at $Q^2=0$. Thus it cannot be found as a 
free particle, though bound states with finite radii can contain it. The 
masses of these bound states would be of the order 
$\Lambda_\mathrm{QCD}$ purely by dimensional considerations. Hence the 
energy scale, or the energy gap, for gluon decoupling is 
$\mathcal{O}(\Lambda_\mathrm{QCD})$. This is soft confinement, meaning 
that the gluons are bound together only by finite-distance dynamics.

 \section{The case of QED}\label{sec_qed}

  Let us now turn our attention to the case of the running coupling in 
QED.

\begin{figure}[ht]{
 \centerline{\epsfig{file=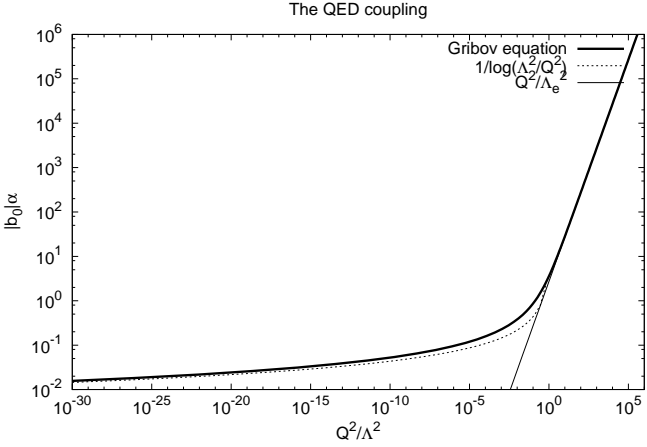, width=9cm}}
 \caption{The QED running coupling calculated using the Gribov equation,
compared against the leading-order perturbative behaviour and the
$\propto Q^2$ behaviour.
 \label{fig_qed_coupling}}}\end{figure}

  We show the plot of eqn.~\ref{eqn_gribov_soln_exp} for negative $b_0$ 
in fig.~\ref{fig_qed_coupling}. This plot is generated by evaluating 
$Q^2/\Lambda^2$ as a function of $b_0\alpha$

  The high-energy behaviour is governed by the $\Gamma=1$ fixed point of 
eqn.~\ref{eqn_beta_gribov}. The large-coupling limit of 
eqn.~\ref{eqn_gribov_soln_exp} is given by:
 \begin{equation}
  \lim_{\alpha\to\infty}\alpha(Q^2)=
  \frac{Q^2}{\left|b_0\right|\Lambda_e^2}.
  \label{eqn_linear_evolution}
 \end{equation}
  This corresponds to the straight line shown in 
fig.~\ref{fig_qed_coupling}. We also show the ordinary one-loop 
perturbative result in the same plot.

  As in QCD, we expect that the value of $\Lambda$ is highly sensitive 
to various factors, including the higher-order corrections. Therefore we 
do not trust the constant of proportionality in 
eqn.~\ref{eqn_linear_evolution} to be accurate. On the other hand, we 
believe that the $\propto Q^2$ behaviour to be correct since, as 
explained earlier, the photon decouples from the theory only in this 
case.

  This $\propto Q^2$ behaviour for the UV coupling yields a 
current--current contact interaction, similar to the NJLVL model 
\cite{njlvl}. The coupling constant, $M_\mathrm{NJLVL}^2$, by 
eqn.~\ref{eqn_linear_evolution}, is given by $4\pi b_0\Lambda_e^2$. The 
interaction Lagrangian density is, for space-like exchange:
 \begin{equation}
  \mathcal{L}_I=\frac1{M_\mathrm{NJLVL}^2}
  \left(\overline\psi\gamma_\mu\psi\right)
  \left(\overline\psi\gamma_\mu\psi\right).
  \label{eqn_njlvl}
 \end{equation}
  Ref.~\cite{njlvl} reports the formation of a massless Goldstone mode.  
This is consistent with Gribov's EWSB mechanism \cite{gribovtop} based 
on top quark condensation \cite{topcondensation} due to the $U(1)_Y$ 
Landau pole. There are other bosonic modes such as the Higgs boson and 
the massive axial vector boson reported in ref.~\cite{njlvl}. Although 
the photon decouples at $Q^2=\Lambda^2$, these bosons remain physical 
even above this scale\footnote{This behaviour of QED at high energy is 
similar to that in media with large screening. When either the coupling 
$\alpha$ or the response of the medium $\left|b_0\right|$ is large, the 
interaction becomes point-like, i.e., the contact-term interaction.
  The oscillation of the medium gives rise to states like the charge 
density wave in solids and the Higgs and Goldstone bosons in high-energy 
QED.}.

  Our results are supported by a numerical study of high-energy QED 
\cite{kogut_dagotto_kocic}, which yielded exactly the same conclusions, 
namely that the photon decouples, yielding a contact interaction which 
gives rise to chiral-symmetry breaking.

  We believe that the effect of gravity does not spoil the applicability 
of this EWSB mechanism, because the relevant gravitational coupling 
remains small above the Planck scale \cite{odagirigrav}. 

  There are no fixed points for time-like running, and therefore 
eqn.~\ref{eqn_ordinary_running} remains a valid description for negative 
$Q^2=-q^2$. In this case, the real part of the coupling changes its sign 
at $q^2=\Lambda^2$, indicating that the photon becomes a ghost, or 
unphysical, above this value of $q^2$.

  Let us now compare our results with eqn.~\ref{eqn_gribov_qed} due to 
Gribov. As mentioned in the introduction, the strong-coupling limit of 
this equation is given by:
 \begin{equation}
  \left(\frac{d^2}{d\xi^2}+2\frac{d}{d\xi}\right)\frac1g=
  -\frac2{g^2}.
  \label{eqn_gribov_qed_strong_coupling}
 \end{equation}
  where $\xi=\ln Q^2/\Lambda^2$.
  If we omit the $g^{-2}$ term, the general solution of this equation 
is:
 \begin{equation}
  g^{-1}=C_1+C_2e^{-2\zeta}=C_1+C_2\Lambda^4/Q^4.
 \end{equation}
  $C_1$ and $C_2$ are the constants of integration. The coupling 
asymptotically tends to $C_1^{-1}$ at high energy.

  With the inclusion of the $g^{-2}$ term, this is no longer constant, 
and we obtain $g^{-1}\to\zeta^{-1}$, or:
 \begin{equation}
  g\to\ln(Q^2/\Lambda^2),
 \end{equation}
  which is the behaviour obtained by Gribov \cite{gribovqed}.

  Let us consider the modification to 
eqn.~\ref{eqn_gribov_qed_strong_coupling} due to the running of the 
photon propagator, viz.\ eqn.~\ref{eqn_internal_renormalization_effect}:
 \begin{equation}
  g_\mathrm{eff}=g(1-\Gamma).
  \label{eqn_defn_geff}
 \end{equation}
  We make this substitution because the $g$ on the right-hand side of 
eqns.\ \ref{eqn_gribov_qed} and \ref{eqn_gribov_qed_strong_coupling} 
arises from the solution of eqn.~\ref{eqn_gribov_a}. By exactly the same 
argument as that in the paragraph following 
eqn.~\ref{eqn_internal_renormalization_effect}, there is an extra 
renormalization effect which multiplies $g$ by $(1-\Gamma)$.

  Since the left-hand side of eqn.~\ref{eqn_gribov_qed_strong_coupling} 
is derived by manipulating the derivatives of $\Pi(q^2)$, it is 
unaffected by this effect.
  We then have:
 \begin{equation}
  \left(\frac{d^2}{d\xi^2}+2\frac{d}{d\xi}\right)\frac1g=
  -\frac2{g^2(1-\Gamma)^2}.
  \label{eqn_gribov_qed_strong_coupling_modified}
 \end{equation}
  We have assumed that the strong-coupling approximation is valid in the 
sense that $g_\mathrm{eff}$ is greater than approximately $1$.
  An asymptotic solution, this time, is:
 \begin{equation}
  \Gamma\to1-\sqrt{2\Lambda^2/Q^2}, \qquad g\to Q^2/\Lambda^2.
 \end{equation}
  This choice of $\Gamma$ makes $g(1-\Gamma)^2\to2$ asymptotically 
constant, and so $g\to Q^2/\Lambda^2=e^\xi$ solves 
eqn.~\ref{eqn_gribov_qed_strong_coupling_modified}.

  Hence we believe that our results are consistent with that of Gribov, 
provided that one takes into account the effect due to photon 
renormalization inside the vacuum-polarization operator.

  Before concluding this section, we would like to mention one property 
of eqn.~\ref{eqn_gribov_soln_exp}, which does not seem particularly 
useful, but we think is worth mentioning.

  Obviously eqn.~\ref{eqn_gribov_soln_exp} cannot be inverted to obtain 
$\alpha$ as some elementary function of $Q^2$, but certain moments of it 
can be evaluated in a closed form and are finite.
  Let $x=Q^2/\Lambda_e^2$ and $a=\left|b_0\right|\alpha$. Then:
 \begin{equation}
  \int_0^\infty x^na^{-m}da=\int_0^\infty e^{-n/a}a^{n-m}da.
 \end{equation}
  Then by the definition of the Euler Gamma function, this becomes:
 \begin{equation}
  n^{1+n-m}\Gamma(m-n-1).
 \end{equation}
  The moments of $x^n$ under $a$ may not be useful at all, but they are 
related to the moments of $a^{-m}$ under $x$, which may seem slightly 
more useful.

 \section{Conclusions}\label{sec_conclusions}

  We derived a local Gribov equation for the gluon/photon Green's 
function $D(q)$, and solved it 
for both QCD and QED.

  Our derivation is based on taking the second derivative of 
$\mathrm{Tr}\ln D(q)$. We separated out the parts due to the running 
coupling from the parts due to the renormalization of the leading, 
two-gluon(photon)-insertion term. Both using Gribov's 
logarithmic-enhancement argument and using a Dyson--Schwinger-type 
expression, the part due to the running coupling is shown also to be of 
the form which corresponds to the double emission of zero-momentum 
gluons/photons, and so the Gribov equation can be written down in a 
compact form.

  The Gribov equation gives an equation for the running of the coupling 
in a closed form. We obtained the solution of this equation in an 
analytical form, for both QCD and QED. Although we do not expect our 
equation to be fully applicable to QED, we argued that it has sensible 
UV behaviour.

  In the case of QCD, we obtain an $\alpha_S$ which is finite, but has a 
branch-point singularity at $Q^2=\Lambda_\mathrm{QCD}^2$.
  We interpret this as being due to the formation of gluonic 
super-critical states, which makes the vacuum unstable. However, 
adopting Gribov's scenario, the singularity is moved off the axis due to 
the decay of the gluon into light-quark super-critical states. 
Furthermore, with the re-organization of the (Dirac-sea) vacuum, these 
decays become forbidden, giving rise to a coupling which only has 
singularities along the time-like axis. However, there are singularities 
on both $+i\varepsilon$ and $-i\varepsilon$ sides of the time-like axis, 
giving rise to, in the low-energy limit, a coupling which is IR-free.
  The gluon is then softly confined.

  In the case of QED, the Gribov equation has both an IR fixed point 
given by $\Gamma=0$ and UV fixed point given by $\Gamma=1$. The latter 
gives a UV $\propto Q^2$ limiting behaviour for the coupling. The photon 
decouples from the theory. The high energy limit of QED is then given by 
the contact interaction. This supports Gribov's scenario of EWSB by 
top-quark condensation due to the strong $U(1)_Y$ interaction near the 
would-be Landau pole.

  We believe that our formalism can be applied also to the problem of 
scale generation in gravity \cite{odagirigrav}.


 \acknowledgement

  Acknowledgements:
  major part of the work was carried out at the Institute of Physics, 
Academia Sinica, Taiwan.
  We are indebted to Profs.~B.R.~Webber and Yu.L.~Dokshitzer for 
discussions and penetrative remarks which helped sharpen our 
understanding.
  The present form of the exposition owes much to the critical remarks 
by the EPJC referee. Much of the derivation and results presented here, 
including some central results, have been worked out upon his/her 
suggestions.

 \end{document}